\documentclass[prl,amsmath,amssymb,superscriptaddress,preprint]{revtex4-1}
\usepackage[utf8]{inputenc}
\usepackage{dcolumn}
\usepackage{bm}
\usepackage{float}
\usepackage{graphicx}
\usepackage[colorlinks]{hyperref}
\usepackage{hyperref}
\usepackage[version=4]{mhchem}
\usepackage{subfigure}

\begin{document}

\title{Linking dynamics and structure in highly asymmetric ionic liquids}

\author{Mariana E. Far\'ias-Anguiano}
\email{email: mfariasa@ifisica.uaslp.mx}
\affiliation{División de Ciencias e Ingenierías, Universidad de Guanajuato, Loma del Bosque 103, 37150 León, Gto., México}
\affiliation{Instituto de F\'isica ``Manuel Sandoval Vallarta'', Universidad 
Aut\'onoma de San Luis Potos\'i, \'Alvaro Obreg\'on 64, 78000 San Luis Potos\'i, SLP, M\'exico.}

\author{Ernesto C. Cort\'es-Morales}
\email{email: ecortesm@nd.edu}
\affiliation{Department of Chemical and Biomolecular Engineering, University of Notre Dame, Notre Dame, IN 46556, USA.}

\author{Jonathan K. Whitmer}
\email{email: jwhitme1@nd.edu}
\affiliation{Department of Chemical and Biomolecular Engineering, University of Notre Dame, Notre Dame, IN 46556, USA.}
\affiliation{Department of Chemistry and Biochemistry, University of Notre Dame, Notre Dame, IN 46556, USA.}

\author{Pedro E. Ram\'irez-Gonz\'alez} 
\email{email: pedro.email@uaslp.mx}
\affiliation{CONACyT-Instituto de F\'isica ``Manuel Sandoval Vallarta'', Universidad 
Aut\'onoma de San Luis Potos\'i, \'Alvaro Obreg\'on 64, 78000 San Luis Potos\'i, SLP, M\'exico.}

\date{\today}

\begin{abstract}
    We explore an idealized theoretical model for the transport of ions within highly asymmetric ionic liquid mixtures. A primitive model (PM)-inspired system serves as a representative for asymmetric ionic materials (such as liquid crystalline salts) which quench to form disordered, partially-arrested phases. Self-Consistent Generalized Langevin Equation (SCGLE) Theory is applied to understand the connection between the size ratio of charge-matched salts and their average mobility. Within this model, we identify novel glassy states where one of the two charged species (either the macro-cation or the micro-anion) are arrested, while the other retains mobility. We discuss how this result is useful in the development of novel single-ion conducting phases in ionic liquid based materials.
\end{abstract}

\maketitle


The design of structurally stable, high conductivity electrolytes is a key materials challenge for modern energy storage, with wide-ranging implications for the future of battery technologies. Ionic liquids (ILs) and their close relatives---polymerized ionic liquids (PILs), room temperature molten salts (RTMS) and deep eutectic solvents (DES)---have been of particular interest, comprising a class of materials attractive for favorable material properties~\citep{Plechkova2008, Abbott2004}, including a wide temperature range over which the materials are chemically stable, in the liquid phase, and maintain high ionic conductivity~\citep{Abbott2004,Ohno2006}. In addition to applications in energy storage\cite{Lewandowski2009}, ionic liquids have been promising as catalysts~\citep{Zhao2002,Olivier2010}, plasticizers\cite{Lu2009} and in desalination membranes~\cite{Guo2021}. 

Many common ILs are formed by combining a larger cation, often affixed with alkyl chains of varying length, with a smaller organic or inorganic anion~\cite{Dong2012}. Many methods have been explored to discern IL structural properties, both theoretically and in simulation. For the later, it is common practice to implement Molecular Dynamics and Monte Carlo simulations\cite{Tong2018} which allow for connections between molecular structure and thermodynamic or dynamical properties. Many developments have been achieved in IL simulations, and for this work we will refer to the specialized studies\cite{Cao2018} into the phase transition of IL crystals\cite{Quevillon2018}. Even with specialized and accurate simulations for the prediction of material properties in a chemical complex of interest, general trends can be missed by such specific models. Hence the need to develop model representations and theoretical tools that aid in guiding the design and study of new IL materials. Despite the fact that many of the described systems have been analyzed in crystalline phase, the search for amorphous glassy states represents new opportunities for such materials. Other scientific communities (such as metallurgists) have found great advantage in the application of amorphous materials.\cite{Axinte2012,Miracle2004} Thus, the study of amorphous solid conductors could be a new branch of development within superionic conductors community.

The basis for past analytical studies of ionic liquid properties is the closure to the Ornstein-Zernike\cite{Ornstein1914} equation formulated by Waisman and Lebowitz, known as the MSA (mean spherical approximation)\citep{Waisman1972-1,Waisman1972-2}. Here we explore a simplified representation of ILs based on the primitive model (PM)\cite{Simonin1996} solution of the MSA with dynamical behavior determined through a generalized Langevin equation representation. We will return to this formulation in detail later, for now it is sufficient to mention that this is not the only theoretical approximation for the study of electrolytes and IL, there are plenty of theoretical approximations such as: the widely used Debye-H\"uckel limiting laws\cite{Weingartner2008}, the Generalized MSA\cite{Stell1975}, or the Binding MSA\cite{Bernard2014}, among several others.

\begin{figure}
    \includegraphics[scale=0.5]{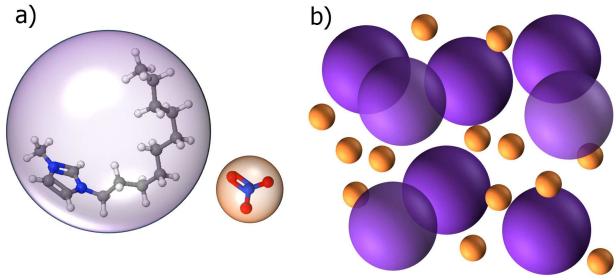}
    \caption{%
        The system that will be analyzed is a binary mixture of charged hard spheres, where a size asymmetry is present in the system. The smaller spheres are able to move freely even when the bigger spheres are all in a glassy state. This model is based on the \ce{[C_{10}Mim]+[NO_3]-} ionic liquid.%
    }
    \label{fig:model}
\end{figure}

The theoretical framework known as Self-Consistent Generalized Langevin Equation (SCGLE) theory has been previously applied to a wide variety of systems, both purely theoretical and realistic model substances. It has been able to successfully predict thermodynamic\cite{Lira-Escobedo2021} and dynamical properties, in particular: dynamical arrest diagrams, MSD and diffusion coefficients of complex chemical compounds such as room temperature molten salts,\cite{Ramirez2016} and low-density colloidal Wigner glasses.\cite{Sanchez-Diaz2009} In the present work, we employ the SCGLE framework to determine the arrest diagram and MSD of a theoretical system composed of a charged hard sphere binary mixture with size asymmetry between the cation and anion. Without loss of generality, we will refer to the larger species (macro-ion) as the cation and the smaller species (micro-ion) as the anion, as depicted in Fig.~\ref{fig:model}. Though similar models have been explored before using the SCGLE, the ionic mobility in the mixture and its effect on conductivity has not been fully explored yet, which is our primary interest here. The use of size asymmetry, even without charge symmetry, results in domains within the temperature-concentration plane which yield arrest of the cation, anion, or both. The partially-arrested states, where one of the ionic components remains fluid, arise in this system due to the size asymmetry, and allow us to map out a phase space relevant for single-ion conducting electrolytes.\citep{Ryu2005,Porcarelli2016} While our model is a coarse approximation of the molecular structures present in real ionic liquids and their derivatives, this work provides a crucial first step towards a detailed theoretical modeling of ionic liquids, and toward engineering materials with optimal conductivity and stability across a wide range of temperatures.

An illustration of the coarse-grained model we utilize as well as a typical configuration of such system is shown in Fig.~\ref{fig:model}. The size of the cation and anion is roughly determined by the ratio of radii of gyration for the ionic liquid \ce{[C10MIm]+[NO3]-}. In the resulting model, we define parameters for the cation using the subscript $\rm p$ reflecting its positive charge, while the anion parameters are labelled by a subscript $\rm m$ reflecting their negative charge. Both species are modeled as charged hard spheres. The diameter of the larger cation species is chosen as the unit of length to normalize our calculations, $\sigma_{\rm p} = 1.0$. The anion is chosen to be ten times smaller, $\sigma_{\rm m} = 0.1$. Both species have an equal and opposite charge $\lvert z_i \rvert = e$, and the total number of ions is such that the system complies with the global electroneutrality condition required in the MSA closure $\sum_i\rho_i z_i = 0$. The size asymmetry in this system is captured by the dimensionless quantity $\xi = \sigma_p / \sigma_m = 10.0$, which has been previously shown to be within the limits of validity for the SCGLE framework.\cite{Juarez-Maldonado2008}

For simplicity in our calculations, we will employ the Hiroike solution\cite{Hiroike1977} to the MSA previously mentioned and proposed by Blum\cite{Blum1975} within the PM: 
\begin{equation}
\label{eq:msa}
\begin{array}{r c l r l}
    {h_{ij}}\left( r \right) & = & - 1 &{\rm{for}}& r \le {\sigma_{ij}}\\
    {c_{ij}}\left( r \right) & = &- \beta \frac{{{e^2}{q_i}{q_j}}}{{{\varepsilon _0}r}}& {\rm{for}}& r >{\sigma _{ij}}
\end{array}
\end{equation}

\noindent for the direct correlation function of the ion pair $c_{ij}(r)$ and the total correlation function $h_{ij}(r)$, with ${\sigma _{ij}} = \left( {{\sigma _i} + {\sigma _j}} \right)/2$ the arithmetic mean of ionic diameters $\sigma_i$ and  $\sigma_j$. These direct correlation functions are used as additional input to the SCGLE representation for a size-asymmetric binary mixture, which was studied previously in Ref.~\cite{Ramirez2016}.
 
A numerical solution to the Ornstein--Zernike equation under the conditions outlined above may be computed for the non-trivial case of size asymmetry. The consequent structure factors $S_{ij}(k)$ are introduced into the SCGLE equilibrium equations in Fourier space to obtain the arrest factor $\gamma_i$ which characterizes the mobility of each ionic species.

\begin{widetext}
\begin{equation}\label{eq:gamma}
  \frac{1}{{{\gamma_i }}} = \frac{1}{{3{{\left( {2\pi } \right)}^3}}}\int {{d^3}k{\mkern 1mu} {k^2}{{\left\{ {\lambda {{\left[ {\lambda  + {k^2}\gamma } \right]}^{ - 1}}} \right\}}_{ii }}}  \nonumber \\
  {\left\{ {c\sqrt n \lambda S{{\left[ {\lambda S + {k^2}\gamma } \right]}^{ - 1}}\sqrt n h} \right\}_{ii }} 
\end{equation}
\end{widetext}

\noindent Here, $\lambda(k)$ is a diagonal matrix with elements $\lambda_{ij} \equiv \delta_{ij}[1 + (k/k_c^{i})^2]^{(-1)}$. The value ${k_c^i = 2\pi(1.305) / \sigma_i}$ is obtained from  the location of the minimum value attained in $S_{ij}(k)$ after its primary maximum. The $c_{ij}(k)$ and $h_{ij}(k)$ are the correlation functions in Eq.~\ref{eq:msa} and the matrix $n$ is defined by $(\sqrt n )_{ij} \equiv \delta_{ij} \sqrt{n_i}$ where $n_i$ is the number density of species i. The complete formulation of Eq. \ref{eq:gamma} has been reported in detail\citep{Juarez-Maldonado2008} by Juarez-Maldonado. What is most relevant here is its physical meaning.

\begin{figure}[ht!]
    \includegraphics[width=0.48\textwidth]{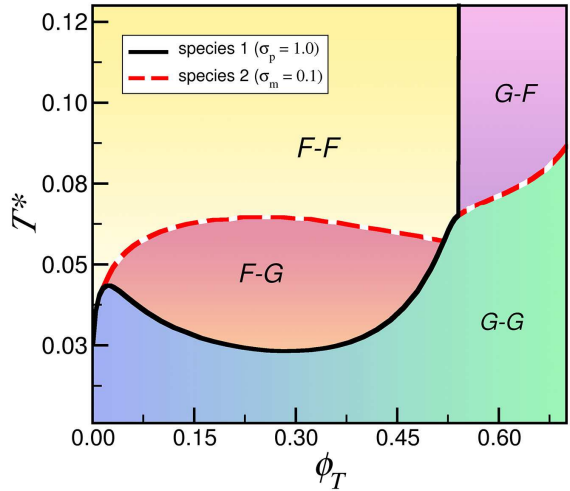}
    \caption{%
        Kinetic arrest diagram as function of the total packing fraction $\phi_T$ and the reduced temperature $T^*$. Four distinct regions are apparent, characterized by the mobility of the cation and anion species (with cation listed first): a completely fluid region (F--F), a completely arrested region (G--G), and two partially-arrested regimes (F--G and G--F). The character of each region is explored further in subsequent plots. %
    }
    \label{fig:phase}
\end{figure}

The parameter $\gamma_i$ has a unique value for each pair of thermodynamic properties in our system, specifically $\phi_{\rm T}$ and $T^*$, where $\phi_T$ is the total volume fraction of our binary hard sphere mixture ($\phi_T = \pi (n_p\sigma_p^3 + n_m\sigma_m^3)/6$) and $T^*= k_BT \sigma_{ij}\varepsilon_0/e^2$ is the reduced temperature. Values of $\gamma_i \rightarrow \infty$ correspond to ergodic fluid states, while finite values of $\gamma_i < 1.0$ define glassy arrested states. By fixing the value of $\phi_{\rm T}$ (which defines an isochore) and computing $\gamma_i$ values for one species within a range of temperatures, it is possible to map the change in values and delimit regions in the phase space which exhibit kinetic arrest of one or both species. The results are shown in Fig.~\ref{fig:phase} where four distinct regions have been found for changes in values of $\gamma_i$ for the cation and anion species. In each ``phase'', the kinetic state (fluid F or glass G) of the cation is listed first and that of the anion is listed second. The lower region corresponds to finite small values of $\gamma_i$ in both species, and is a completely arrested phase. At higher temperatures, adhesive interactions between opposite charges are suppressed, so that in systems which have low enough volume fraction to avoid jamming both species have fluid-like mobility. More interesting is the regions at top right (high volume fraction, high temperature) where only one of the ionic species is arrested. The partially-arrested F-G and G-F regions will be the focus of our subsequent analysis.

\begin{table*}[ht!]
\centering
\begin{tabular}{ c | c | c | c | c }
 \hline

    Point & Packing Fraction ($\phi_T$) & Reduced Temperature ($T^*$) & Phase ($+$ ion) & Phase ($-$ ion) \\ 
 \hline\hline
 1 & 0.40 & 0.080 & F & F\\[0.7ex] 
 2 & 0.40 & 0.033 & F & G\\[0.7ex] 
 3 & 0.40 & 0.009 & G & G\\[0.7ex] 
 4 & 0.53 & 0.062 & \multicolumn{2}{c}{Intersection} \\[1ex] 
 5 & 0.60 & 0.080 & G & F\\[0.7ex] 
 6 & 0.60 & 0.060 & G & G\\[1ex] 
 \hline
\end{tabular}
\caption{Representative points within the arrest diagram (Fig.~\ref{fig:phase}) explored in subsequent analyses.}
\label{table:1}
\end{table*}

\begin{figure*}[ht!]
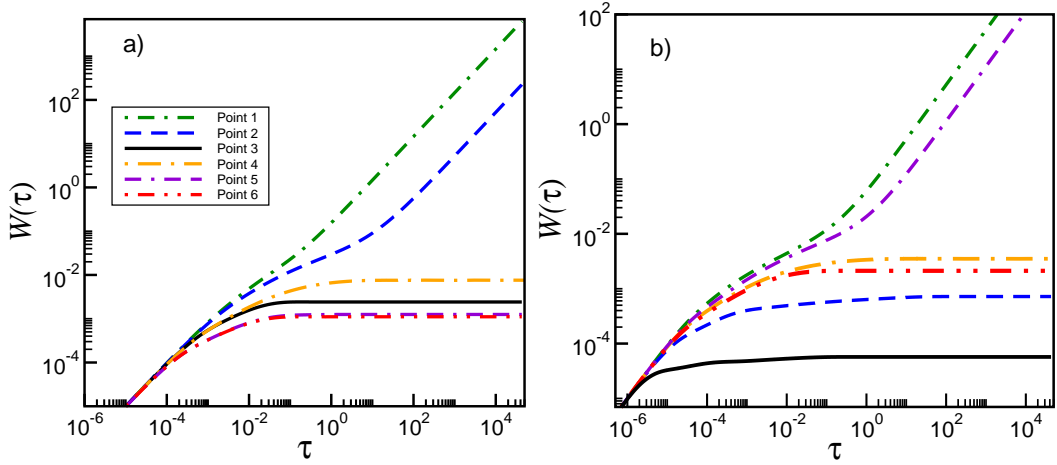

\includegraphics[width=0.42\textwidth]{plots/msd1.eps}
\includegraphics[width=0.42\textwidth]{plots/msd2.eps}
\caption{MSD ($W$) plotted as a function of non-dimensionalized time $\tau$.  results for seven points in the arrest diagram corresponding to different regions for the species with a) $\sigma_p = 1.0$ and b) $\sigma_m = 0.1$. The locations of these points in the ($\phi_T,T^*$) plane is given in Table~\ref{table:1}}
  \label{fig:msd}
\end{figure*}

To collect further insights into the phase transition phenomena being observed, we will select six points in the arrest diagram, as shown in table \ref{table:1}, each point in a different region, one point in the intersection of the four regions, and we will calculate the MSD of the two ionic species at each region using the dynamic version of the SCGLE theory. This framework has been extensively explored and specifics of its implementation and analysis are presented in detail elsewhere \citep{Yeomans-Reyna2001,Chavez-Rojo2005,Yeomans-Reyna2007}. The dynamic version of the SCGLE framework requires the same input as before, which is the structure factor for each ionic species, and unique values of total volume fraction $\phi_T$ and reduced temperature $T^*$. Several dynamic quantities are obtained in the solution of the coupled SCGLE equations, one of which is the MSD. The results of our calculations are shown in Fig.~\ref{fig:msd}, where panel (a) exhibits the cation behavior and panel (b) the anion behavior. The evolution in MSDs can be observed spanning several regimes in time, with fluid behaviors observed as curves asymptotically approaching linear growth in time as $t\rightarrow\infty$, and arrested behaviors approaching a constant value after diffusive motion at short time-scales.

It is worth noting that the MSD behavior at the two possible partially-arrested states, labelled by points 2 (blue dashed line) and 5 (purple dot-dashed line) in Table~\ref{table:1} and Fig.~\ref{fig:msd}, exchanges the identity of the mobile and immobile species; above $\phi_T=0.52$ the partially-arrested states have mobile small ions with the large ions arrested, below this value the partially-arrested states have mobile large ions with the small ions arrested. Long-time arrested states follow the same trend as the relaxation time when the slope goes to zero is the same.

\begin{figure*}[ht!]
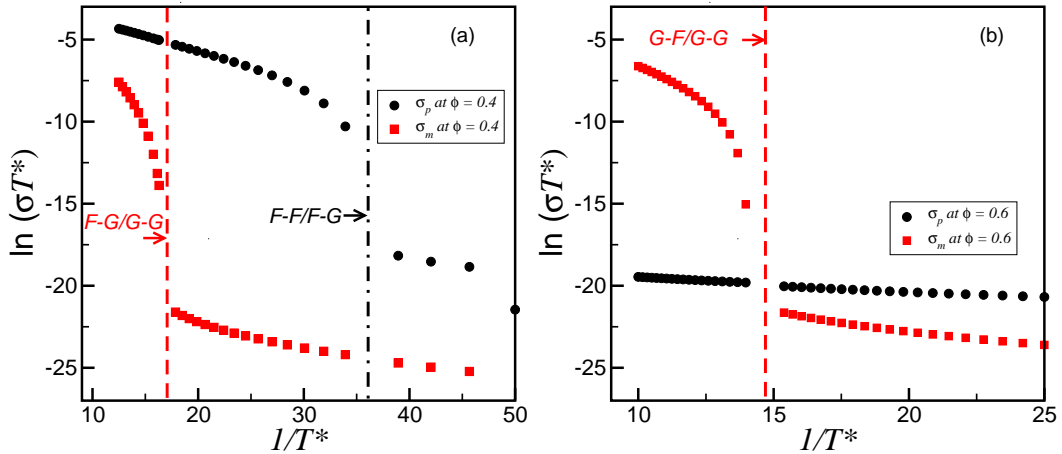

\includegraphics[width=0.42\textwidth]{plots/conduct_phi4.eps}
\includegraphics[width=0.42\textwidth]{plots/conduct_phi6.eps}
\caption{Qualitative results for the prediction of the electrical conductivity as function of $1/T^*$. a) Isochoric trajectory at $\phi_T = 0.4$, we go through three different phase regions and the cations retain their high conductivity at various temperatures. b) For the isochoric trajectory at $\phi_T = 0.6$, going through two phase regions and this time, the anions are responsible for most of the conductive behaviour.}
  \label{fig:cond}
\end{figure*}

We can utilize our solutions to extract the electrical conductivity, $\sigma \approx D_L$ for each ionic species, where $D_L$ is the long-time diffusion coefficient \cite{Roling2001}. In Fig.~\ref{fig:cond}, the logarithm of the conductivity times the reduced temperature as function of the inverse of the reduced temperature is calculated for both species at a total volume fraction a) $\phi_T = 0.4$, and b) $\phi_T = 0.6$. An intriguing behavior arises here, similar to previously reported studies, \citep{Ramirez2016,Sanz2008}, where partially-arrested regions emerge with the larger ionic species diffusing among the smaller ionic species. This is a counter-intuitive behavior, though this phenomenon can be explained by a less efficient charge screening of the larger ions from the smaller ones, due to volume effects, and thus the small ion--small ion electrostatic interaction is effectively stronger than the opposite charged interaction, to the point where it promotes the formation of a glass phase of smaller ions while large ions remain fluid. Thus, the electronic conductivity is possible due to the anomalous diffusion of only one species of larger ions in a glassy state of smaller ions. 

\begin{figure*}
    \includegraphics[scale=0.65]{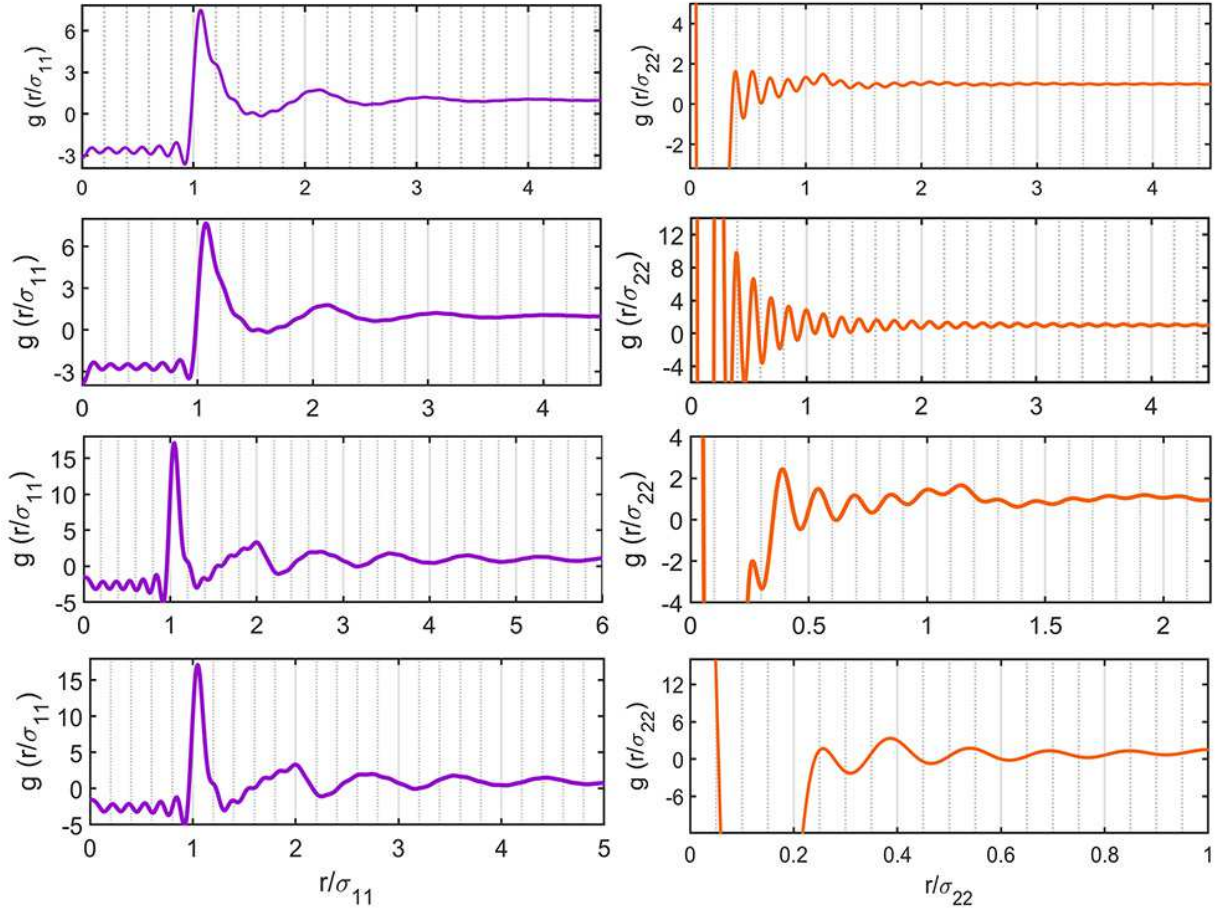}
    \caption{%
        Radial distribution functions used as inputs in the theoretical analysis. Points 1, 2, 5 and 6 from table 1 were selected and the $g(r)$ for the species 1 is presented with purple line, while the ones for species 2 are shown in orange lines.%
    }
    \label{fig:rdfs}
\end{figure*}

Figure \ref{fig:rdfs} shows the RDF calculated numerically from the inverse fast Fourier transform (IFFT) of the Hiroike structure factors, with the purple lines corresponding to the cation and orange lines to the anion. The 4 pairs correspond with points 1, 2, 5 and 6 and are used as a visual corroboration of the predicted arrest diagram \ref{fig:phase}, as anionic RDFs in points 2 and 4 seem ordered and agree with a glass phase, while cationic RDFs at points 1 and 2 are similar and effectively correspond to fluid phases.

For practical applications, however, we are interested in the explicit case where the cations become arrested but anions remain fluid, as at Point 5 in Table~\ref{table:1}. Since charges may be exchanged in our model without changing any of the thermodynamic and dynamic behavior, this configuration is a representative state for ionic liquid crystals such as those comprising the polymers in Ref.~\cite{Merrill2020} with small lithium cations and bulky TFSI-derived anions. Our results suggest bulky cations can be immobilized by steric hindrance (or other molecular interactions) leaving the smaller species free to move through the electrolyte, as evidenced both by the MSD results in Fig.~\ref{fig:msd} and the conductivity results in Fig.~\ref{fig:cond}(b). This presents an intriguing design strategy for single-ion conductors where the salt species is designed to maximize the glassiness of the macroionic species at concentrations and temperatures where the microion remains mobile.

In this work, a theoretical binary mixture of charged hard spheres is analyzed in the framework of the Self Consistent Generalized Langevin Equation Theory (SCGLE). In the equilibrium version of the SCGLE, the $\gamma_i$ parameter is used to determine limit lines where arrest states for each $i$ species exist. A point in this phase space is characterized by a unique coordinate pair of $\phi_T$ and $T^*$ parameters, the total volume fraction and the reduced temperature, respectively. This enables determination of a kinetic arrest diagram for a given size asymmetry of charged particles via the dynamic SCGLE theory. Results here are in agreement with prior equilibrium results, but are able to elucidate partially-arrest regions that may be exploited for their novel charge transport behavior. These regions occur for considerable size asymmetries between cation and anion (such as the 10:1 ratio explored here), and are anticipated to be relevant for systems such as ionic liquid crystals and polymerized ionic liquids, where interstitial conductive domains arise between arrested cations. These results demonstrate new and significant behaviors relevant for developing energy storage technologies, and offer predictions which may be realized by engineering new molecular structures and materials.

PERG appreciates the assistance of J. Lim\'on Castillo and  acknowledge LANIMFE for the infrastructure provided during this project. MFA and PERG acknowledge the financial support of CONACyT through grants: C\' atedras CONACyT No. 1631
and CB-2015-01 No. 257636. ECCM and JKW acknowledge support from the University of Notre Dame through an International Collaboration Grant.

\bibliography{asymsize}

\end{document}